\documentclass[12pt,letterpaper]{article}
\usepackage{fullpage}
\usepackage{citesort}
\usepackage{graphicx}
\usepackage{amsmath}
\usepackage{amssymb}
\usepackage{bm} 
\usepackage{tikz}
\usetikzlibrary{snakes}

\allowdisplaybreaks[1]
\def\be#1\ee{\begin{equation}#1\end{equation}}
\def\bal#1\eal{\begin{align}#1\end{align}}
\def\bmu#1\emu{\begin{multline}#1\end{multline}}
\def\bga#1\ega{\begin{gather}#1\end{gather}}
\newcommand{\ba}{\begin{array}}
\newcommand{\ea}{\end{array}}
\newcommand{\n}{\notag}

\renewcommand{\b}{\mathbf}
\renewcommand{\d}{\partial}

\newcommand{\scs}{\scriptstyle}

\title{\textbf{Functional approaches to infrared \\
Yang-Mills theory in the Coulomb gauge}}

\author{A Weber$^{1,}$\thanks{Speaker. Email: 
\texttt{axel@ifm.umich.mx}}\;, M Leder$^2$, 
J M Pawlowski$^{3,4}$ and H Reinhardt$^2$ \\[2mm]
\and
\normalsize $^1$ Instituto de F\'isica y Matem\'aticas, Universidad Michoacana 
de San Nicol\'as de Hidalgo, \\
\normalsize Edificio C-3, Ciudad Universitaria, 58040 Morelia,
Michoac\'an, Mexico 
\and
\normalsize $^2$ Institut f\"ur Theoretische Physik, Universit\"at T\"ubingen, 
Auf der Morgenstelle 14, \\
\normalsize 72076 T\"ubingen, Germany
\and
\normalsize $^3$ Institut f\"ur Theoretische Physik, Universit\"at Heidelberg, 
Philosophenweg 16, \\
\normalsize 69120 Heidelberg, Germany 
\and
\normalsize $^4$ ExtreMe Matter Institute EMMI, GSI Helmholtzzentrum f\"ur 
Schwerionenforschung, \\
\normalsize Planckstr.\ 1, 64291 Darmstadt, Germany}

\date{June 10, 2011}

\begin{document}

\maketitle

\begin{abstract}
We present the current status of ongoing efforts to use functional methods,
Dyson-Schwinger equations and functional renormalization group
equations, for the description of the infrared regime of nonabelian (pure)
gauge theories in the Coulomb gauge. In particular, we present a new 
determination of the color-Coulomb potential with the help of the functional 
renormalization group that results in an almost linearly rising potential 
between static color charges at large spatial distances.
\end{abstract}

\newpage
\section{Introduction}

Important progress has been achieved over the last decade in the description
of the deep infrared region of nonabelian gauge theories with the help of
functional methods, employing Coulomb gauge fixing. By functional methods
we refer to semi-analytical tools that do \emph{not} make use of the
discretization of space-time as does lattice gauge theory. Specifically,
equations of Dyson-Schwinger type arising from a variational principle have
been used, and more recently functional renormalization group equations.
In this contribution, we will report on the current status of these
investigations. We will focus exclusively on pure gauge theories, more
specifically SU(N) Yang-Mills theory, but include static color charges
so as to obtain a description of the heavy quark potential, as in
quenched approximations.

We will start by briefly describing the general theoretical setup: the
Hamiltonian framework is used, where the Weyl \emph{and} Coulomb gauge 
conditions, $A_0^a (\b{x}) = 0$ and $\nabla \cdot \b{A}^a (\b{x}) = 0$, are
imposed on the SU(N) gauge fields. Physical states are described by
wave functionals of $\b{A}^a (\b{x})$ with scalar product
  \be
  \langle \phi | \psi \rangle = \int D[\b{A}] \, J[\b{A}] \, 
  \phi^\ast [\b{A}] \, \psi [\b{A}] \,. \label{scalarproduct}
  \ee
Here, $J[\b{A}]$ stands for the Faddeev-Popov (FP) determinant $J[\b{A}] = 
\text{Det} \, (-\nabla \cdot \b{D})$ with the spatial covariant derivative
$\b{D}^{ab} = \delta^{ab} \nabla + g f^{abc} \b{A}^c (\b{x})$. The
functional integral in \eqref{scalarproduct} is understood to be restricted
to spatially transverse gauge fields, i.e., to those that fulfill the gauge
fixing conditions.

The dynamics is defined by the Christ-Lee Hamiltonian $H$ \cite{CL80} that
we do not write out. In the presence of a static color charge density
$\rho^a_q (\b{x})$, $H$ contains the interaction term
  \be
  H_q = \frac{1}{2} \int d^3 x \, d^3 y \, \rho^a_q (\b{x}) 
  F^{ab} (\b{x}, \b{y}) \rho^b_q (\b{y}) 
  \ee
with the integral kernel
  \be
  F^{ab} (\b{x}, \b{y}) = \langle \b{x}, a | (-\nabla \cdot 
  \b{D})^{-1} (-\nabla^2) (-\nabla \cdot \b{D})^{-1} 
  | \b{y}, b \rangle \,.
  \ee
The vacuum expectation value 
$\left\langle F^{ab} (\b{x}, \b{y}) \right\rangle$ is
called the color-Coulomb potential. It is suppposed to give the dominant
contribution to the confining interaction between color charges. More
precisely, for large spatial distances the color-Coulomb potential provides 
an upper bound for the Wilson potential \cite{Zwa98}.

For the following, it will be convenient to write the FP
determinant in a local form by introducing ghost fields, 
  \be
  J [\b{A}] = \text{Det} \, (-\nabla \cdot \b{D}) = \int D 
  [\bar{c}, c] \, \exp \left(-\int d^3 x \, \bar{c}^a (\b{x}) 
  (-\nabla \cdot \b{D}^{ab}) c^b (\b{x}) \right) \,.
  \ee
In our analysis, we will focus on the equal-time correlation functions, i.e.
the vacuum expectation values of products of the field operators 
$\b{A}^a (\b{x})$ (transverse), $c^a (\b{x})$ and $\bar{c}^a (\bf{x})$. 
We can easily write down an expression for the generating functional of these 
correlation functions,
  \bal
  Z [\b{J}, \eta, \bar{\eta}] &= \int D[\bar{c}, c, \b{A}] \,
  e^{-\int d^3 x \, \bar{c} (-\nabla \cdot \b{D}) c} \left| \psi 
  [\b{A}] \right|^2 \n \\
  &\phantom{=} {}\times \exp \left(\int d^3 x \left[ \b{J}^a 
  (\b{x}) \cdot \b{A}^a (\b{x}) + \bar{c}^a (\b{x}) 
  \eta^a (\b{x}) + \bar{\eta}^a (\b{x}) c^a (\b{x}) \right] 
  \right) \,, \label{generatingfunctional}
  \eal
where $\psi [\b{A}]$ is the vacuum wave functional.
If we formally define an ``action'' $S[\b{A}]$ through
$\left| \psi [\b{A}] \right|^2 = e^{-S[\b{A}]}$, 
\eqref{generatingfunctional} looks like the usual generating functional of
Euclidean Green's functions in the covariant Lagrangian formulation of the
theory, only in three instead of four dimensions. Of course, $S[\b{A}]$
is a complicated and a priori unknown functional of $\b{A}^a (\b{x})$. We
will parametrize the ``propagators'', the equal-time two-point correlation
functions of the theory, in the most general way (restricted by symmetries)
as follows:
  \bga
  \left\langle A^a_i (\b{p}) \, A^b_j (-\b{q}) \right\rangle =
  \frac{1}{2 \omega (p)} \, \delta^{ab} \left( \delta_{ij} -
  \frac{p_i p_j}{p^2} \right) (2 \pi)^3 \delta (\b{p} - \b{q}) \,, \\
  \left\langle c^a (\b{p}) \, \bar{c}^b (-\b{q}) \right\rangle =
  \left\langle \langle \b{p}, a | (-\nabla \cdot \b{D})^{-1} 
  | \b{q}, b \rangle \right\rangle =
  \frac{d (p)}{p^2} \, \delta^{ab} \, (2 \pi)^3 \delta 
  (\b{p} - \b{q}) \,. \label{ghostpropagator}
  \ega
Here and in the following, we use the notation $p = |\b{p}|$. The
functions $\omega (p)$ and $d (p)$ will be of central interest in the rest
of this contribution. Notice that the ghost propagator \eqref{ghostpropagator}
is just the vacuum expectation value of the inverse FP operator
(or rather, its integral kernel).

\section{Variational principle: Dyson-Schwinger equations}

A set of equations of Dyson-Schwinger type for the equal-time correlation 
functions of the theory was obtained in Ref.\ \cite{SS01}
from the variational principle, using a Gaussian ansatz for the vacuum 
wave functional. The contribution of the FP determinant 
was fully taken into account in \cite{FR04,RF05}. We write the ansatz for the 
vacuum functional as
 \be
 \left| \psi [\b{A}] \right|^2 = e^{-\widetilde{S}[\b{A}]} \,, \qquad
 \widetilde{S} [\b{A}] = \frac{1}{2} \int \frac{d^3 p}{(2\pi)^3} \, 
 A^a_i (-\b{p}) \, 2 \widetilde{\omega} (p) \, A^a_i (\b{p}) \,.
 \label{gaussian}
 \ee
Then the variational principle with respect to the unknown function
$\widetilde{\omega} (p)$,
 \be
 \frac{\delta}{\delta \widetilde{\omega} (p)} \, \langle H \rangle = 0 \,,
 \ee
leads to a gap equation for the equal-time gluon propagator. The detailed
form of the equation as well as the approximations involved in its
derivation can be found in \cite{FR04,RF05}. 

The gap equation involves, apart from the gluon propagator, the ghost
propagator and the color-Coulomb potential, hence further input is needed
in order to arrive at a closed system of equations. The generating functional
\eqref{generatingfunctional} can be used to derive a Dyson-Schwinger (DS)
equation for the ghost propagator in the usual way:
 \be
  p^2  d^{-1} (p) \equiv \Big(
  \raisebox{-4pt}{\parbox{1.4cm}{\begin{center}
  \begin{tikzpicture}[>=stealth,scale=1.4]
   \draw[dash pattern=on 1.5pt off 1.2pt] (0.2,0) -- (1,0) %
node[below] {$\scs p$};
  \fill (0.6,0) circle (1.7pt);
   \draw[->,very thin] (0.33,0) -- (0.3,0);
   \draw[->,very thin] (0.83,0) -- (0.8,0);
  \end{tikzpicture}
  \end{center}}} \Big)^{-1}
  = \, Z_c p^2 \,
  - \parbox{3.1cm}{\begin{center}
  \begin{tikzpicture}[>=stealth,scale=1.4]
   \begin{scope}[snake=coil, segment length=3pt, segment amplitude=2pt]
   \foreach \x in {-5,5,...,165}
     \draw[snake] [xshift=0.5cm] (\x+19:0.5cm) -- (\x:0.5cm);
   \end{scope}
   \draw[dash pattern=on 1.5pt off 1.2pt] (-0.4,0) -- (1.4,0) %
node[below] {$\scs p$};
   \draw[->,very thin] (1.18,0) -- (1.15,0);
   \draw[->,very thin] (0.77,0) -- (0.74,0);
   \draw[->,very thin] (0.27,0) -- (0.24,0);
   \draw[->,very thin] (-0.22,0) -- (-0.25,0);
   \fill (0.5,0) circle (1.6pt);
   \fill (0.5,0.5) circle (1.65pt);
   \fill (0,0) circle (1pt);
   \fill (1,0) circle (1pt);
  \end{tikzpicture} 
  \end{center}} \,. \label{diagDSghost}
 \ee
In the diagrams, we represent the full equal-time ghost propagator by a
dashed line and the gluon propagator by a curly line, with a dot on the lines.
By extending Taylor's non-renormalization theorem \cite{Tay71} to the present 
situation, we have replaced in \eqref{diagDSghost} the full ghost-gluon vertex
(one of the vertices on the right-hand side) with the bare one. This
replacement is also used in the gap equation.
 
For the color-Coulomb potential, 
 \be
  \left\langle F^{ab} (\b{p}, -\b{q}) \right\rangle =
  \left\langle \langle \b{p}, a | (-\nabla \cdot \b{D})^{-1} (-\nabla^2)
  (-\nabla \cdot \b{D})^{-1} | \b{q}, b \rangle \right\rangle =
  V_c (p) \, \delta^{ab} \, (2 \pi)^3 \delta (\b{p} - \b{q}) \,,
 \ee
we use the following parameterization and diagrammatic representation 
motivated by the appearance of the inverse FP operator 
[cf.\ \eqref{ghostpropagator}]:
 \be
  V_c (p) = \frac{d(p)}{p^2} \, p^2 f(p) \, \frac{d(p)}{p^2}
  = \raisebox{-4pt}{\parbox{2.5cm}{\begin{center}
  \begin{tikzpicture}[>=stealth,scale=1.4]
   \draw[dash pattern=on 1.5pt off 1.2pt] (0.2,0) -- (1.6,0) %
node[below] {$\scs p$};
   \fill (0.6,0) circle (1.7pt);
   \filldraw[fill=white] (0.9,0) circle (1.8pt);
   \fill (1.2,0) circle (1.7pt);
   \draw[->,very thin] (0.4,0) -- (0.37,0);
   \draw[->,very thin] (1.43,0) -- (1.4,0);
  \end{tikzpicture}
  \end{center}}} \,,
 \ee
thereby defining the Coulomb form factor $f(p)$. Of course, the function
$f(p)$ is itself unknown, and before discussing the possibility of
determining it in terms of the gluon and ghost propagators in the next
section, we will resort to the factorization hypothesis \cite{Zwa04}
 \be
 \left\langle (-\nabla \cdot \b{D})^{-1} 
 (-\nabla^2) (-\nabla \cdot \b{D})^{-1} \right\rangle
 = \left\langle (-\nabla \cdot \b{D})^{-1} \right\rangle (-\nabla^2) 
 \left\langle (-\nabla \cdot \b{D})^{-1} \right\rangle \,,
 \ee
which is equivalent to
 \be
 V_c (p) = \frac{d(p)}{p^2} \, p^2 \, \frac{d(p)}{p^2} \,, 
 \label{factorization}
 \ee
or $f(p) = 1$. Adopting this (so far unjustified) assumption, we obtain a 
closed system of equations.

Before discussing the numerical solutions of the equations, 
we have to comment on the
Gribov-Zwanziger confinement scenario \cite{Gri78,Zwa92}. In brief, the
idea is that the existence of Gribov copies (gauge-equivalent but different
transverse gauge field configurations) forces one to restrict the functional
integral over the gauge field to the first Gribov region where the FP operator 
is positive definite.\footnote{Actually, the first Gribov region contains
Gribov copies itself, and it is necessary to further restrict the integral
to the so-called fundamental modular region. It is likely that this
additional complication does not affect the following argument (in the
case of the Coulomb gauge).} By a 
statistical argument, in the infrared (IR) regime the dominant contribution to
the functional integral comes from the region close to the Gribov
horizon where the FP operator has zero modes. Since the
ghost propagator is the vacuum expectation value of the inverse FP operator,
it may be argued that the ghost propagator $d(p)/p^2$ should be more singular 
in the IR than $p^{-2}$, thus $d^{-1} (p=0) = 0$, the ``horizon
condition'' \cite{Zwa91}. Hence, one should look for solutions that fulfill
the horizon condition.

It turns out that there are two different solutions of this type 
\cite{FR04,ERS07}.\footnote{Actually, in these numerical solutions a
more sophisticated approximation of the Coulomb form factor, going
beyond the factorization hypothesis, was used (see below).} 
Both show scaling behavior in the IR, i.e.,
the equal-time propagators obey power laws in this kinematical regime.
Furthermore, among the different contributions to the gluon propagator
in the gap equation, the ghost loop diagram completely dominates the IR
behavior, a property known as ghost dominance. These facts make it possible 
to even obtain analytical solutions for the propagators in the 
IR region \cite{Zwa04,SLR06}. With the notations
 \be
 \omega (p) = A \, p^{-\alpha} \,, \qquad d(p) = B \, p^{-\beta} \,,
 \ee
one obtains in this way a general sum rule for the IR exponents:
 \be
 \alpha = 2 \beta - 1 \,. \label{sumrule}
 \ee
Consistent solutions exist for the values
$(\alpha = 0{.}592, \beta = 0{.}796)$ and $(\alpha = 1, \beta = 1)$. 
One may also define a running coupling constant from the ghost-gluon vertex
as
\be
\alpha(p) = \frac{8}{3} \, \frac{g^2(p)}{4\pi} \,, \qquad
  g^2(p) = g_B^2 \, \frac{p}{\omega(p)} \, d^2 (p) \label{runningcoupling}
\ee
($g_B$ is the bare coupling constant). In the ultraviolet (UV), the 
solutions show asymptotic freedom (although not with the correct power of
$\ln p$ due to the approximations to the gap equation), while $\alpha(p)$
saturates at a constant value in the IR. Analytically, one obtains
$N_c \, \alpha(0) = 11{.}99$ for the solution with $\beta = 0{.}796$, and
$N_c \, \alpha(0) = 16 \pi/3$ for $\beta = 1$ \cite{SLR06}.

\begin{figure}
\begin{center}
\includegraphics[width=0.5\textwidth,clip]%
{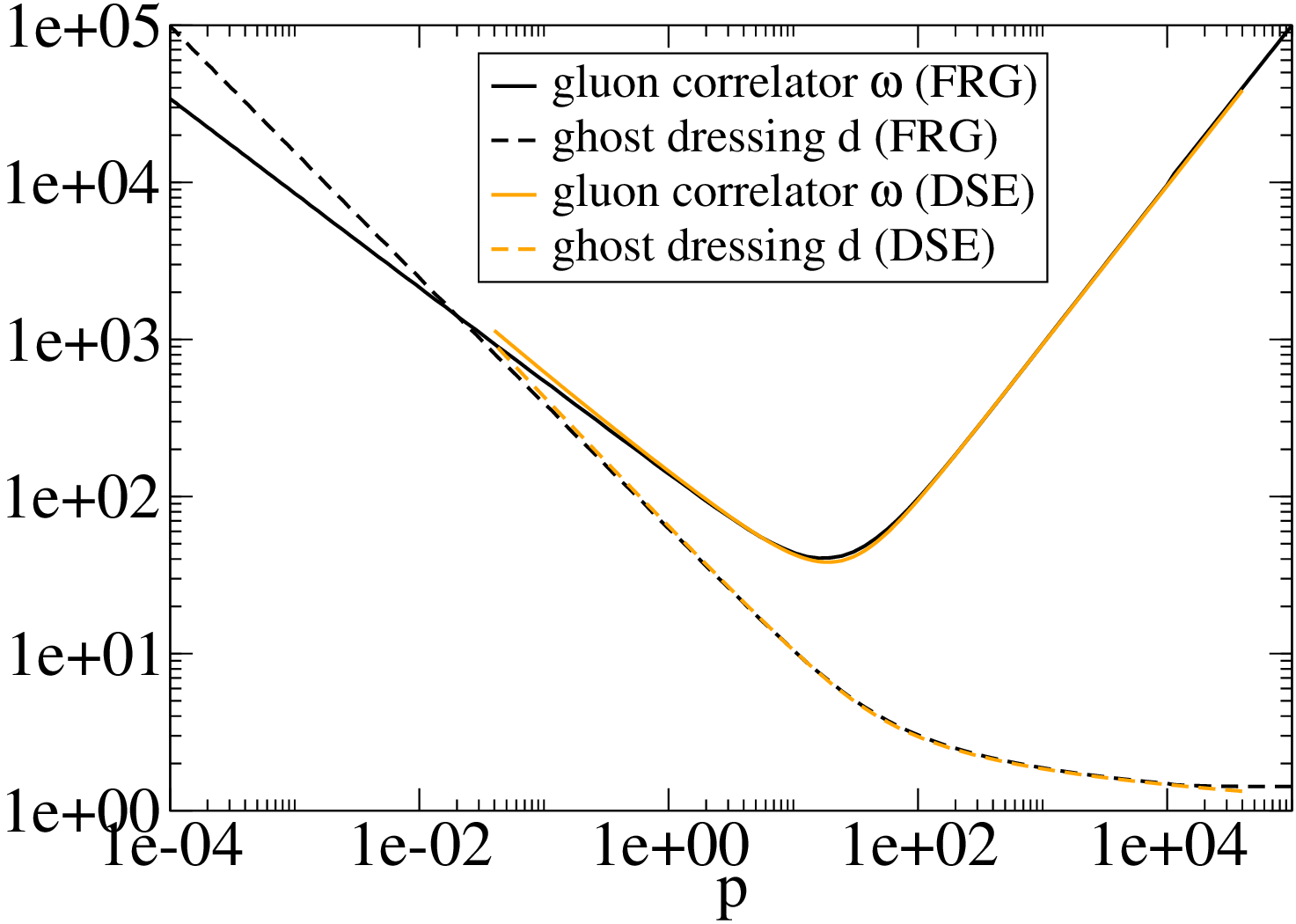}
\end{center} 
\vspace{-4mm}
\caption{\label{figcomparison}Comparison of the results for the propagators 
from the two different sets of equations \cite{LPR10}. The label ``DSE'' stands
for the set that contains the gap equation and ``FRG'' for the set that uses
\eqref{diagDSgluon}.}
\end{figure}

To close this section, we comment on a possible drastic simplification of 
the equations: if one uses, instead of the general Gaussian ansatz 
\eqref{gaussian}, the lowest-order perturbative vacuum wave functional
 \be
 |\psi [\b{A}]|^2 = e^{-S_0 [\b{A}]} = \exp \left( -\frac{1}{2} \int 
 \frac{d^3 p}{(2\pi)^3} \,  A^a_i (-\b{p}) \, 2 p \, A^a_i (\b{p}) \right) 
 \label{perturbativevacuum}
 \ee 
in the generating functional \eqref{generatingfunctional}, the complicated 
gap equation may be replaced by the following DS equation for 
the gluon propagator:
 \be
  2 \omega (p) \equiv \Big(
  \raisebox{-4pt}{\parbox{1,4cm}{\begin{center}
  \begin{tikzpicture}[scale=1.4]
   \begin{scope}[snake=coil, segment length=3pt, segment amplitude=2pt,%
   line before snake=1.5pt]
   \draw[snake] (0.2,0) -- (1,0) node[below] {$\scs p$};
   \end{scope}
   \fill (0.58,0) circle (1.75pt);
  \end{tikzpicture}
  \end{center}}} \Big)^{-1}
  = \, 2 Z_A p \,
  - \parbox{3.1cm}{\begin{center}
  \begin{tikzpicture}[>=stealth,scale=1.4]
   \begin{scope}[snake=coil, segment length=3pt, segment amplitude=2pt,%
   line before snake=1.5pt]
   \draw[snake] (-0.45,0) -- (0.05,0);
   \draw[snake] (0.98,0) -- (1.44,0) node[below] {$\scs p$};
   \end{scope}
   \draw[dash pattern=on 1.5pt off 1.2pt] (0.5,0) circle (0.5cm);
   \draw[->,very thin] (0.85,-0.35) -- (0.88,-0.32);
   \draw[->,very thin] (0.15,-0.35) -- (0.18,-0.38);
   \draw[->,very thin] (0.85,0.35) -- (0.82,0.38);
   \draw[->,very thin] (0.15,0.35) -- (0.12,0.32);
   \fill (0.5,0.5) circle (1.6pt);
   \fill (0.5,-0.5) circle (1.6pt);
   \fill (0,0) circle (1pt);
   \fill (1,0) circle (1pt);
  \end{tikzpicture}
  \end{center}} \,. \label{diagDSgluon}
 \ee
In particular, \eqref{diagDSgluon} does
not make use of the factorization hypothesis. Due to ghost dominance,
the gap equation approaches \eqref{diagDSgluon} in the IR. Somewhat
surprisingly, we have found numerically that the solutions obtained with the
two different sets of equations coincide over the whole momentum range from
the IR to the UV to good numerical precision (for $\beta = 0{.}796$) 
\cite{LPR10}. In Fig.\ \ref{figcomparison}, ghost and gluon propagators are 
represented in a double-logarithmic plot for the two different sets of 
equations. 
The slight discrepancy in the IR is due to the lower numerical precision of the
earlier calculation (``DSE'') in \cite{FR04}.

\section{Color-Coulomb potential and factorization hypothesis}

The color-Coulomb potential is more directly related to physically
observable quantities than the gluon and ghost propagators. For the
solution with $\beta = 1$, the factorization hypothesis \eqref{factorization}
leads to $V_c (p) \propto p^{-2 - 2 \beta} = p^{-4}$ in the IR which
corresponds to a potential in position space that rises exactly linearly
for large distances. Unfortunately, the approximation used in the UV [see
our remark following \eqref{runningcoupling}] does not permit to relate the 
(Coulomb) string tension to the scale $\Lambda_{\text{QCD}}$.

We will now turn to the question of whether the factorization hypothesis 
is actually justified. To this end, it is convenient to represent the 
color-Coulomb potential with the help of a composite operator $K$,
  \bga
  \left\langle \langle \b{x}, a | (-\nabla \cdot \b{D})^{-1}
  (-\nabla^2) (-\nabla \cdot \b{D})^{-1} | \b{y}, b \rangle \right\rangle
  = \left\langle c^a (\b{x}) K \bar{c}^b (\b{y}) \right\rangle_{\text{GI}} 
  \,, \n \\
  K = \int d^3 z \, \bar{c}^d (\b{z})
  (-\nabla^2_\b{z}) c^d (\b{z}) \,, \label{compositeoperator}
  \ega
where the index GI (gluon-irreducible) on the vacuum expectation value
means that 
one has to restrict the contributing diagrams to those where the operator $K$ 
remains connected to the external points when all gluon lines are cut. The
Coulomb form factor $f(p)$ is then precisely the form factor of the
composite operator $K$. Introducing $K$ in the standard way in the generating
functional \eqref{generatingfunctional}, one may derive a DS
equation for $f(p)$. After suitable approximations, one arrives at
(see also \cite{FR04})
 \be
  p^2 f (p) \equiv 
  \raisebox{-4pt}{\parbox{1.4cm}{\begin{center}
  \begin{tikzpicture}[>=stealth,scale=1.4]
   \draw[dash pattern=on 1.5pt off 1.2pt] (0.2,0) -- (1,0) %
node[below] {$\scs p$};
   \filldraw[fill=white] (0.6,0) circle (1.9pt);
   \draw[->,very thin] (0.33,0) -- (0.3,0);
   \draw[->,very thin] (0.83,0) -- (0.8,0);
  \end{tikzpicture}
  \end{center}}}
  = \, Z_f p^2 \,
  + \parbox{3.1cm}{\begin{center}
  \begin{tikzpicture}[>=stealth,scale=1.4]
   \begin{scope}[snake=coil, segment length=3pt, segment amplitude=2pt]
   \foreach \x in {-5,5,...,165}
     \draw[snake] [xshift=0.5cm] (\x+19:0.5cm) -- (\x:0.5cm);
   \end{scope}
   \draw[dash pattern=on 1.5pt off 1.2pt] (-0.4,0) -- (1.4,0) %
node[below] {$\scs p$};
   \draw[->,very thin] (1.18,0) -- (1.15,0);
   \draw[->,very thin] (-0.22,0) -- (-0.25,0);
   \filldraw[fill=white] (0.5,0) circle (1.8pt);
   \fill (0.5,0.5) circle (1.65pt);
   \fill (0.25,0) circle (1.6pt);
   \fill (0.75,0) circle (1.6pt);
   \fill (0,0) circle (1pt);
   \fill (1,0) circle (1pt);
  \end{tikzpicture}
  \end{center}} \,. \label{diagDSformfactor}
 \ee

Now \eqref{diagDSformfactor} can be used to close the system of equations
instead of invoking the factorization hypothesis. The result is disappointing:
no solution that fulfills the horizon condition could be found neither
numerically nor analytically \cite{ERSS08}. Numerically, solutions of the
complete system of equations are found to exist only for 
$d^{-1} (0) \gtrsim 0{.}02$. For the latter solutions, $f(p)$ tends to a
constant for $p \to 0$, so that $V_c (p) \propto p^{-2}$ and the
color-Coulomb potential is not confining. We remark
that this negative result is not due to the appearance of $f(p)$ in the
gap equation: in the simplified version that replaces the gap equation
with \eqref{diagDSgluon}, the two DS equations \eqref{diagDSghost}
and \eqref{diagDSgluon} decouple from \eqref{diagDSformfactor}, but still
no solution can be found that would satisfy the horizon condition.
We can now also explain the approximation that was used to obtain
the numerical solutions \cite{FR04,ERS07} (instead of the factorization
hypothesis): the Coulomb form factor was determined from 
\eqref{diagDSformfactor} with the ghost propagators in the loop diagram on 
the right-hand side replaced by the tree-level propagators. The result is a 
form factor that tends toward a constant in the IR.

Let us briefly comment on the most recent results for the equal-time
two-point correlation functions obtained in calculations on space-time lattices
in the Coulomb gauge. They are still somewhat controversial. 
For the ghost propagator,
a UV-behavior $d(p) \propto (\ln p)^{-0{.}33(1)}$ was found in \cite{NVI09}
which is in reasonable agreement with the (also recent) result of 
one-loop perturbation theory $d(p) \propto (\ln p)^{-0{.}36}$ \cite{CWR10}.
On the other hand, the IR-behavior was determined in \cite{NVI09} to be
$d(p) \propto p^{-0{.}435(6)}$ which does not correspond to any of the two
solutions of the DS equations. In \cite{BQR09}, an IR-behavior 
$\omega(p) \propto p^{-1}$ of the gluon propagator was found, which 
is consistent with the $(\beta = 1)$-solution. 
For the UV-behavior of the gluon propagator, however, contradictory results 
have been reported: \cite{BQR09} finds $\omega(p) \propto p$, as opposed to
$\omega(p) \propto p^{1{.}40(2)}$ in \cite{NVI09}. Both results are at odds
with the prediction of one-loop perturbation theory, $\omega(p) \propto p \,
(\ln p)^{0{.}27}$  \cite{CWR10}. As for the color-Coulomb potential,
although the recent results for the UV-behavior do not compare
favorably with perturbation theory and the IR-behavior is inconclusive,
it seems clear that the factorization hypothesis is violated \cite{VIM08}.

\section{The functional renormalization group}

We will now turn to a different functional method, the functional
(or Wilsonian) renormalization group. In order to adapt it to the case
at hand, one starts with the generating functional \eqref{generatingfunctional}
and introduces an IR cutoff $k$ in the following way \cite{LPR10}:
  \bmu
  Z_k [\b{J}, \eta, \bar{\eta}] = \int D[\bar{c}, c, \b{A}] \,
  \exp \left( -\int \frac{d^3 p}{(2 \pi)^3} \, \bar{c}^a (-\b{p})
  R^c_k (p) \, c^a (\b{p}) \right) \\
  {}\times \exp \left( -\frac{1}{2} \int \frac{d^3 p}{(2 \pi)^3} 
  \, A_i^a (-\b{p}) R_k^A (p) \, A_i^a (\b{p}) \right) 
  e^{-\int d^3 x \, \bar{c} (-\nabla \cdot \b{D}) c}
  \, |\psi [\b{A}]|^2 \,
  e^{\, \int d^3 x \, [\b{J} \cdot \b{A} + \bar{c} \eta + 
  \bar{\eta} c]} \,. \label{IRcutoff}
  \emu
The cutoff functions $R_k^c (p)$ and $R_k^A (p)$ have the
properties
 \be
 R_k^{c,A} (p) \to \infty \quad \text{for} \quad p \ll k \,, \qquad
 R_k^{c,A} (p) \to 0 \quad \text{for} \quad p \gg k \,.
 \ee
This means that the IR modes $p \ll k$ in the functional integral 
\eqref{IRcutoff} are heavily suppressed, while in the limit $k \to 0$,
$R_k^{c,A} (p) \to 0$ and $Z_k [\b{J}, \eta, \bar{\eta}]$ tends toward the
full generating functional $Z [\b{J}, \eta, \bar{\eta}]$. In the actual
calculations, we have used an exponential suppression of the IR modes,
  \be
  R^c_k (p) = p^2 r_k (p) \,, \qquad R^A_k (p) = 2 p \, r_k (p) \,, \qquad
  r_k (p) = \exp \left( \frac{k^2}{p^2} - \frac{p^2}{k^2} \right) \,.
  \ee
From \eqref{IRcutoff}, flow equations for the $k$-dependent equal-time 
correlation functions can be derived in the standard way \cite{Wet93}.
They read for the propagators
 \bal
  2 \, \d_k \omega_k (p) \equiv \d_k \bigg[ \Big(
  \raisebox{-4pt}{\parbox{1.4cm}{\begin{center}
  \begin{tikzpicture}[scale=1.4]
   \begin{scope}[snake=coil, segment length=3pt, segment amplitude=2pt,%
   line before snake=1.5pt]
   \draw[snake] (0.2,0) -- (1,0) node[below] {$\scs p$};
   \end{scope}
   \fill (0.58,0) circle (1.75pt);
  \end{tikzpicture}
  \end{center}}} \Big)^{-1}
  - R_k^A (p) \bigg] 
  &= \parbox{3.1cm}{\begin{center}
  \begin{tikzpicture}[>=stealth,scale=1.4]
   \begin{scope}[snake=coil, segment length=3pt, segment amplitude=2pt,%
   line before snake=1.5pt]
   \draw[snake] (-0.45,0) -- (0.05,0);
   \draw[snake] (0.98,0) -- (1.44,0) node[below] {$\scs p$};
   \end{scope}
   \draw[dash pattern=on 1.5pt off 1.2pt] (0.5,0) circle (0.5cm);
   \draw[->,very thin] (0.85,-0.35) -- (0.88,-0.32);
   \draw[->,very thin] (0.15,-0.35) -- (0.18,-0.38);
   \fill (0.15,0.35) circle (1.6pt);
   \fill (0.85,0.35) circle (1.6pt);
   \fill (0.5,-0.5) circle (1.6pt);
   \filldraw[fill=white] (0.5,0.5) circle (2pt);
   \draw (0.45,0.45) -- (0.55,0.55);
   \draw (0.45,0.55) -- (0.55,0.45);
   \fill (0,0) circle (1pt);
   \fill (1,0) circle (1pt);
  \end{tikzpicture}
  \end{center}}
  + \parbox{3.1cm}{\begin{center}
  \begin{tikzpicture}[>=stealth,scale=1.4]
   \begin{scope}[snake=coil, segment length=3pt, segment amplitude=2pt,%
   line before snake=1.5pt]
   \draw[snake] (-0.45,0) -- (0.05,0);
   \draw[snake] (0.98,0) -- (1.44,0) node[below] {$\scs p$};
   \end{scope}
   \draw[dash pattern=on 1.5pt off 1.2pt] (0.5,0) circle (0.5cm);
   \draw[->,very thin] (0.85,0.35) -- (0.82,0.38);
   \draw[->,very thin] (0.15,0.35) -- (0.12,0.32);
   \fill (0.15,-0.35) circle (1.6pt);
   \fill (0.85,-0.35) circle (1.6pt);
   \fill (0.5,0.5) circle (1.6pt);
   \filldraw[fill=white] (0.5,-0.5) circle (2pt);
   \draw (0.45,-0.45) -- (0.55,-0.55);
   \draw (0.45,-0.55) -- (0.55,-0.45);
   \fill (0,0) circle (1pt);
   \fill (1,0) circle (1pt);
  \end{tikzpicture}
  \end{center}} \,, \label{diagflowgluon} \\
  p^2 \d_k d_k^{-1} (p) \equiv \d_k \bigg[ \Big(
  \raisebox{-4pt}{\parbox{1.4cm}{\begin{center}
  \begin{tikzpicture}[>=stealth,scale=1.4]
   \draw[dash pattern=on 1.5pt off 1.2pt] (0.2,0) -- (1,0) %
node[below] {$\scs p$};
  \fill (0.6,0) circle (1.7pt);
   \draw[->,very thin] (0.33,0) -- (0.3,0);
   \draw[->,very thin] (0.83,0) -- (0.8,0);
  \end{tikzpicture}
  \end{center}}} \Big)^{-1} - R^c_k (p) \bigg]
  &= \parbox{3.1cm}{\begin{center}
  \begin{tikzpicture}[>=stealth,scale=1.4]
   \begin{scope}[snake=coil, segment length=3pt, segment amplitude=2pt]
   \foreach \x in {-5,5,...,165}
     \draw[snake] [xshift=0.5cm] (\x+19:0.5cm) -- (\x:0.5cm);
   \end{scope}
   \draw[dash pattern=on 1.5pt off 1.2pt] (-0.4,0) -- (1.4,0) %
node[below] {$\scs p$};
   \draw[->,very thin] (1.18,0) -- (1.15,0);
   \draw[->,very thin] (0.73,0) -- (0.7,0);
   \draw[->,very thin] (0.23,0) -- (0.2,0);
   \draw[->,very thin] (-0.22,0) -- (-0.25,0);
   \fill (0.5,0) circle (1.6pt);
   \fill (0.15,0.35) circle (1.65pt);
   \fill (0.85,0.35) circle (1.65pt);
   \filldraw[fill=white] (0.5,0.5) circle (2pt);
   \draw (0.45,0.45) -- (0.55,0.55);
   \draw (0.45,0.55) -- (0.55,0.45);
   \fill (0,0) circle (1pt);
   \fill (1,0) circle (1pt);
  \end{tikzpicture}
  \end{center}}
  + \parbox{3.1cm}{\begin{center}
  \begin{tikzpicture}[>=stealth,scale=1.4]
   \begin{scope}[snake=coil, segment length=3pt, segment amplitude=2pt]
   \foreach \x in {-5,5,...,165}
     \draw[snake] [xshift=0.5cm] (\x+19:0.5cm) -- (\x:0.5cm);
   \end{scope}
   \draw[dash pattern=on 1.5pt off 1.2pt] (-0.4,0) -- (1.4,0) %
node[below] {$\scs p$};
   \draw[->,very thin] (1.18,0) -- (1.15,0);
   \draw[->,very thin] (-0.22,0) -- (-0.25,0);
   \fill (0.5,0.5) circle (1.65pt);
   \fill (0.77,0) circle (1.6pt);
   \fill (0.23,0) circle (1.6pt);
   \filldraw[fill=white] (0.5,0) circle (2pt);
   \draw (0.45,-0.05) -- (0.55,0.05);
   \draw (0.45,0.05) -- (0.55,-0.05);
   \fill (0,0) circle (1pt);
   \fill (1,0) circle (1pt);
  \end{tikzpicture}
  \end{center}} \,. \label{diagflowghost}
 \eal
Here, the symbol
$\parbox{0.4cm}{\begin{center}
  \begin{tikzpicture}[scale=1.4]
   \filldraw[fill=white] (0.5,0) circle (2pt);
   \draw (0.45,-0.05) -- (0.55,0.05);
   \draw (0.45,0.05) -- (0.55,-0.05);
  \end{tikzpicture}
  \end{center}}$
stands for the insertion of $\d_k R^{c,A}_k$. The non-renormalization theorem
for the ghost-gluon vertex has been used in both equations. Furthermore,
we have neglected diagrams that involve three- and four-gluon vertices,
which is justified for the description of the IR regime if ghost dominance
is assumed. Finally, we have omitted tadpole diagrams in order to be able
to close the system of differential equations. Partial inclusion of the
tadpole diagrams is argued in \cite{LPR10} to lead back to the
DS equations \eqref{diagDSghost} and \eqref{diagDSgluon}
(after integrating over $k$).

\begin{figure}
\begin{center}
\includegraphics[width=0.48\textwidth,clip]{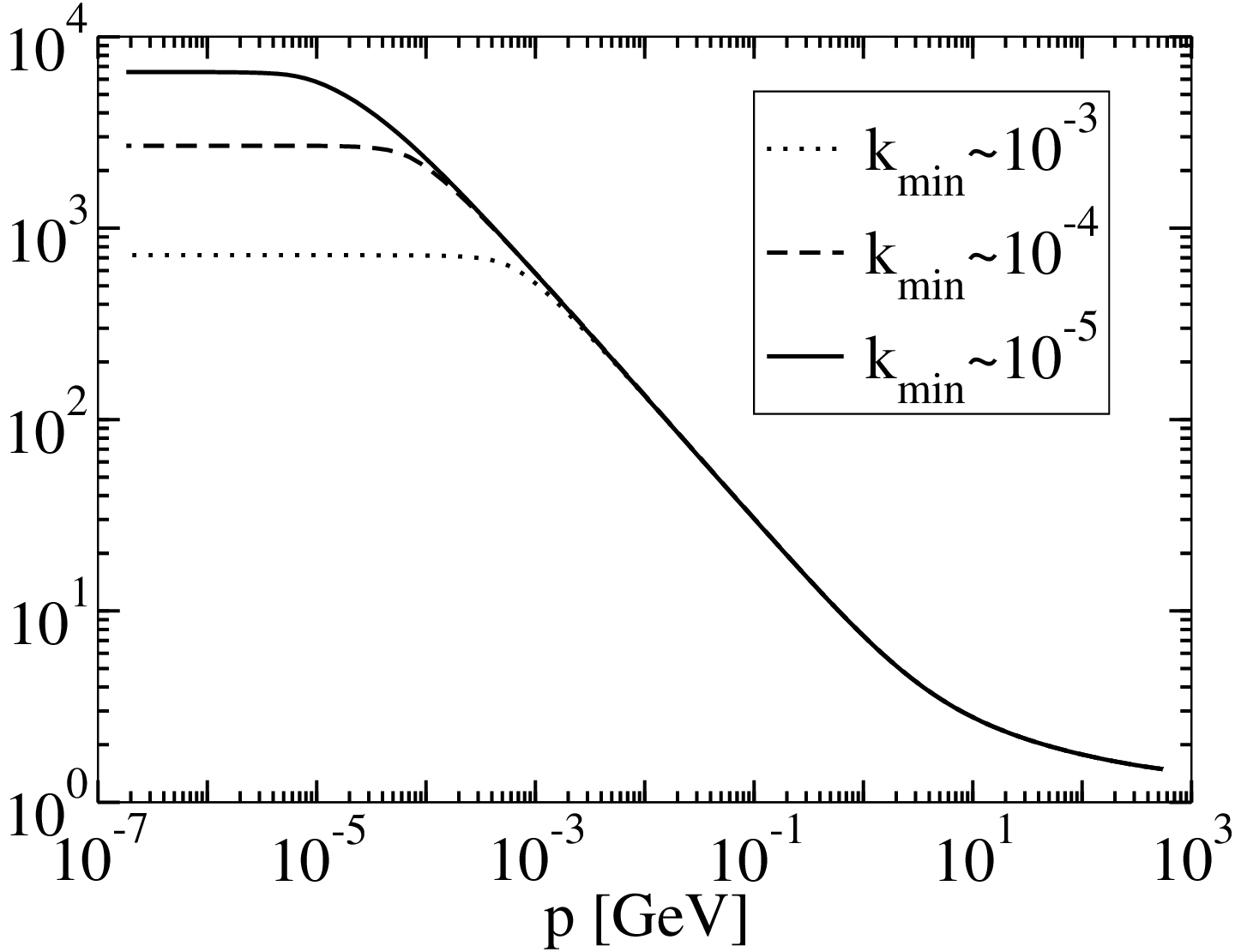}
\hfill
\includegraphics[width=0.48\textwidth,clip]{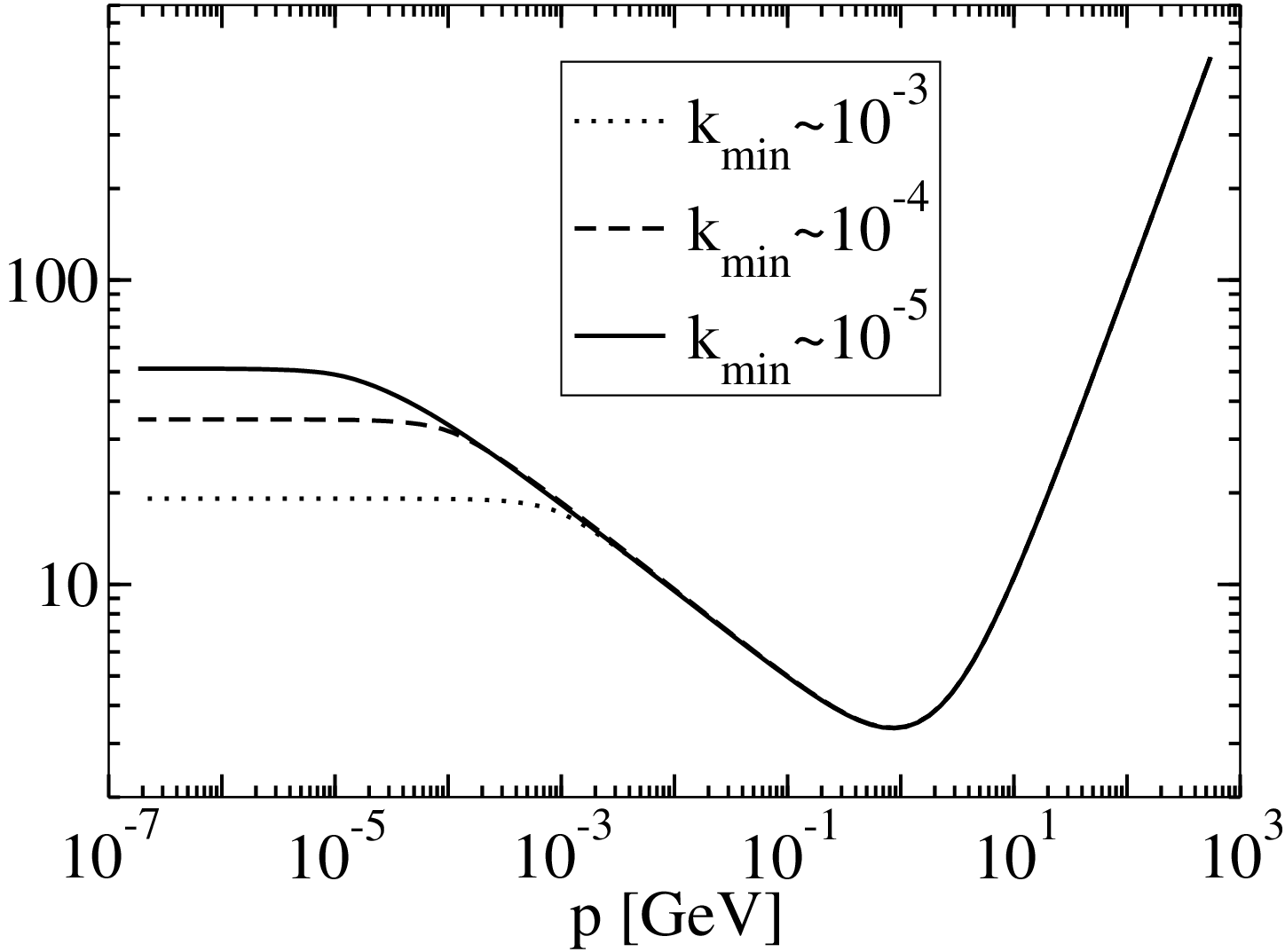}
\end{center}
\vspace{-4mm}
\caption{\label{figpropagators}The results for $d(p)$ (left) and
$\omega (p)$ (right) from the flow equations, for three different values
of $k_{\text{min}}$ \cite{LPR10}.}
\end{figure}

The general strategy is to start integrating the flow equations at a large
value of $k$ where due to asymptotic freedom the ``action'' $S [\b{A}]$
can be replaced with $S_0 [\b{A}]$ from \eqref{perturbativevacuum} and the 
coupling constant is small, so that the initial values of the flow are known.
Then the flow equations are numerically integrated toward $k = 0$, where
$\omega (p) = \omega_{k=0} (p)$ and $d(p) = d_{k=0} (p)$ are read off.
Technically, it is important to convert the differential equations
\eqref{diagflowgluon}, \eqref{diagflowghost} to integral equations first, 
so that the horizon condition and a normalization condition for $\omega (p)$ 
can be conveniently incorporated. The results are presented in Fig.\
\ref{figpropagators}, again as double-logarithmic plots.
For technical reasons, the integration of the flow equations stops at a
minimum value $k_{\text{min}} > 0$. Then $\omega (p) = 
\omega_{k=k_{\text{min}}} (p)$ for $p \gg k_{\text{min}}$, and similarly for
$d(p)$. From Fig.\ \ref{figpropagators} it is clear that the power-law
behavior of the propagators extends toward smaller momenta $p$ as 
$k_{\text{min}}$ is lowered.

The exponents found numerically are $(\alpha = 0{.}28, \beta = 0{.}64)$,
smaller than for both solutions of the DS equations. They obey
the sum rule \eqref{sumrule}. The fact that the exponents come out smaller
than the ones from the DS equations is not entirely unexpected,
since a similar behavior was found in analogous calculations in the Landau
gauge \cite{PLN04}. Generally, the results for the exponents will slightly vary
with the choice of the cutoff functions due to the approximations made in the
system of flow equations. An ``optimized'' choice is expected to give
exponents identical to the ones from the DS equations 
\cite{LPR10}. For the running coupling constant \eqref{runningcoupling}
one also finds saturation in the IR at a slightly smaller value than for
the DS solutions.

\begin{figure}
\begin{center}
\includegraphics[width=0.5\textwidth,clip]{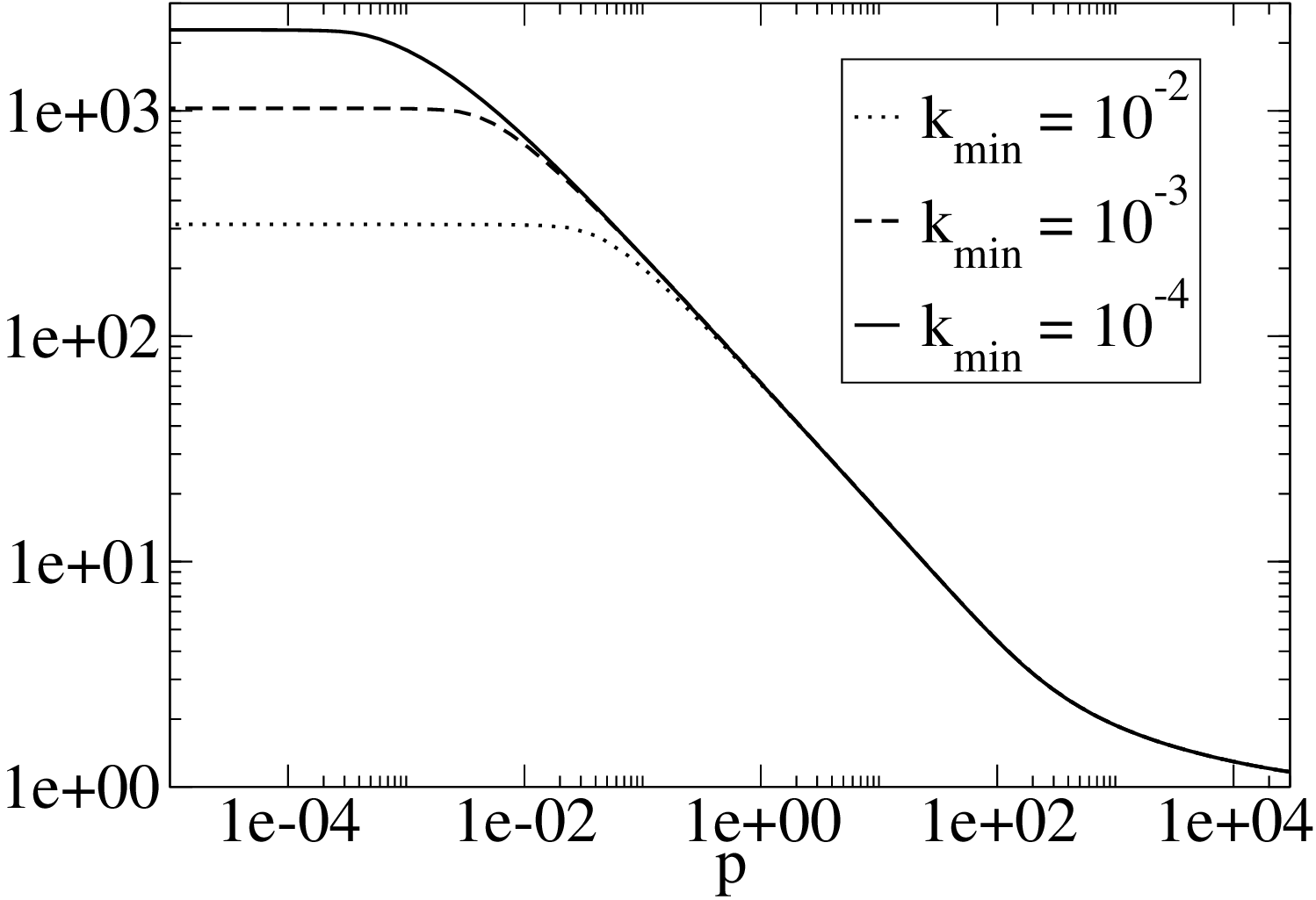}
\end{center}
\vspace{-4mm}
\caption{\label{figformfactor}The Coulomb form factor $f(p)$ from
\eqref{diagflowformfactor}.}
\end{figure}

By incorporating the composite operator $K$ in the functional integral
\eqref{IRcutoff}, one derives (after suitable approximations) the
following flow equation for the Coulomb form factor:
 \be
  p^2 \d_k f_k (p) \equiv \d_k \Big(
  \raisebox{-4pt}{\parbox{1.4cm}{\begin{center}
  \begin{tikzpicture}[>=stealth,scale=1.4]
   \draw[dash pattern=on 1.5pt off 1.2pt] (0.2,0) -- (1,0) %
node[below] {$\scs p$};
   \filldraw[fill=white] (0.6,0) circle (1.9pt);
   \draw[->,very thin] (0.33,0) -- (0.3,0);
   \draw[->,very thin] (0.83,0) -- (0.8,0);
  \end{tikzpicture}
  \end{center}}} \Big)
  = - \parbox{2.9cm}{\begin{center}
  \begin{tikzpicture}[>=stealth,scale=1.4]
   \begin{scope}[snake=coil, segment length=3pt, segment amplitude=2pt]
   \foreach \x in {-5,5,...,165}
     \draw[snake] [xshift=0.5cm] (\x+19:0.5cm) -- (\x:0.5cm);
   \end{scope}
   \draw[dash pattern=on 1.5pt off 0.9pt] (-0.4,0) -- (1.4,0) %
node[below] {$\scs p$};
   \draw[->,very thin] (1.18,0) -- (1.15,0);
   \draw[->,very thin] (-0.22,0) -- (-0.25,0);
   \filldraw[fill=white] (0.5,0) circle (1.8pt);
   \fill (0.15,0.35) circle (1.65pt);
   \fill (0.85,0.35) circle (1.65pt);
   \filldraw[fill=white] (0.5,0.5) circle (2pt);
   \draw (0.45,0.45) -- (0.55,0.55);
   \draw (0.45,0.55) -- (0.55,0.45);
   \fill (0.25,0) circle (1.6pt);
   \fill (0.75,0) circle (1.6pt);
   \fill (0,0) circle (1pt);
   \fill (1,0) circle (1pt);
  \end{tikzpicture}
  \end{center}}
  - \parbox{2.9cm}{\begin{center}
  \begin{tikzpicture}[>=stealth,scale=1.4]
   \begin{scope}[snake=coil, segment length=3pt, segment amplitude=2pt]
   \foreach \x in {-5,5,...,165}
     \draw[snake] [xshift=0.5cm] (\x+19:0.5cm) -- (\x:0.5cm);
   \end{scope}
   \draw[dash pattern=on 1.5pt off 0.9pt] (-0.4,0) -- (1.4,0) %
node[below] {$\scs p$};
   \draw[->,very thin] (1.18,0) -- (1.15,0);
   \draw[->,very thin] (-0.22,0) -- (-0.25,0);
   \fill (0.5,0.5) circle (1.65pt);
   \fill (0,0) circle (1pt);
   \fill (0.15,0) circle (1.6pt);
   \filldraw[fill=white] (0.32,0) circle (2pt);
   \draw (0.27,-0.05) -- (0.37,0.05);
   \draw (0.27,0.05) -- (0.37,-0.05);
   \fill (0.49,0) circle (1.6 pt);
   \filldraw[fill=white] (0.66,0) circle (1.75pt);
   \fill (0.83,0) circle (1.6pt);
   \fill (1,0) circle (1pt);
  \end{tikzpicture}
  \end{center}}
  - \parbox{2.9cm}{\begin{center}
  \begin{tikzpicture}[>=stealth,scale=1.4]
   \begin{scope}[snake=coil, segment length=3pt, segment amplitude=2pt]
   \foreach \x in {-5,5,...,165}
     \draw[snake] [xshift=0.5cm] (\x+19:0.5cm) -- (\x:0.5cm);
   \end{scope}
   \draw[dash pattern=on 1.5pt off 0.9pt] (-0.4,0) -- (1.4,0) %
node[below] {$\scs p$};
   \draw[->,very thin] (1.18,0) -- (1.15,0);
   \draw[->,very thin] (-0.22,0) -- (-0.25,0);
   \fill (0.5,0.5) circle (1.65pt);
   \fill (0,0) circle (1pt);
   \fill (0.17,0) circle (1.6pt);
   \filldraw[fill=white] (0.34,0) circle (1.75pt);
   \fill (0.51,0) circle (1.6pt);
   \filldraw[fill=white] (0.68,0) circle (2pt);
   \draw (0.63,-0.05) -- (0.73,0.05);
   \draw (0.63,0.05) -- (0.73,-0.05);
   \fill (0.85,0) circle (1.6pt);
   \fill (1,0) circle (1pt);
  \end{tikzpicture}
  \end{center}} \,. \label{diagflowformfactor}
 \ee
Making use of the results for $\omega_k (p)$ and $d_k (p)$, this equation
can be integrated. Contrary to the DS equations,
\eqref{diagflowformfactor} has a solution that is represented in Fig.\ 
\ref{figformfactor}.
The IR-behavior is determined numerically to 
\be
 f(p) \propto p^{-\gamma} \,, \qquad \gamma = 0{.}57 \,.
\ee
In particular, of course, $f(p) \neq 1$, and the factorization hypothesis is
violated. With the values for the exponents from the flow equations one obtains
in the IR
  \be
  V_c (p) = \frac{d (p)}{p^2} \, p^2 f(p) \, \frac{d (p)}{p^2} 
  = \frac{1}{p^{2 + 2\beta + \gamma}} = \frac{1}{p^{3{.}85}} \,,
  \ee
close to $V_c (p) \propto p^{-4}$ which would correspond to a linearly rising
potential in position space.

In summary, we find that functional methods are a
powerful tool for the description of the nonperturbative infrared regime
of nonabelian gauge theories. The formulation of these theories in the
Coulomb gauge is particularly convenient, mainly because it gives
direct access to the color-Coulomb potential. The Gribov-Zwanziger
confinement scenario provides a conceptual framework to understand
the confinement mechanism. It can be conveniently implemented via the
horizon condition. In particular, we have seen that an almost linearly
rising color-Coulomb potential is obtained from the functional
renormalization group equations (and the factorization hypothesis
is violated). It has also become clear that the approximations employed still 
have to be improved in order to achieve a quantitatively reliable description
of the infrared region.

\subsection*{Acknowledgments}

A.W. gratefully acknowledges support by CIC-UMSNH.~M.L. and H.R. were
supported by DFG-Re856/6-3.

\end{document}